\documentclass[aip,jap,twocolumn,12pt,reprint]{revtex4-1}

\usepackage{amssymb}
\usepackage{amsmath}
\usepackage{graphicx}
\usepackage{dcolumn}
\usepackage{bm}

\begin{document}

\title{Magnetic cloaking by a paramagnet/superconductor cylindrical tube in the critical state}
\author{S.V.~Yampolskii}
\email[Electronic address: ]{yampolsk@mm.tu-darmstadt.de}
\affiliation{Institut f\"{u}r Materialwissenschaft, Technische Universit\"{a}t Darmstadt,
Jovanka-Bontschits-Stra{\ss}e 2, D-64287 Darmstadt, Germany}
\author{Y.A.~Genenko}
\affiliation{Institut f\"{u}r Materialwissenschaft, Technische Universit\"{a}t Darmstadt,
Jovanka-Bontschits-Stra{\ss}e 2, D-64287 Darmstadt, Germany}
\date{\today }

\begin{abstract}
Cloaking of static magnetic fields by a finite thickness type-II superconductor tube being in the full critical state and surrounded by a 
coaxial paramagnet shell is studied. On the basis of exact solutions to the Maxwell equations, 
it is shown that, additionally to previous studies assuming the Meissner state of the superconductor constituent, 
perfect cloaking is still realizable at fields higher than the field of full flux penetration into the superconductor and
for arbitrary geometrical parameters of both constituents. 
It is also proven that simultaneously the structure is fully undetectable under the cloaking conditions. 
Differently from the case of the Meissner state the cloaking properties in the application relevant critical state are realized, 
however, only at a certain field magnitude.
\end{abstract}

\maketitle

A great research activity has been focused last time on electromagnetic metamaterials 
which exhibit many unique features, providing particularly cloaking of electromagnetic 
waves as well as of low frequency and static magnetic or electric fields~\cite{review2012,Science1,Science2,Science3,Pendry2007,NM2013,ElCloak,NM2008}.
A magnetic cloak is expected to retain undistorted the external field 
outside the cloak, thus being "invisible" for external observation,  and, as far as possible, 
to protect its inner area from the external field penetration. To fulfill these requirements, different cloak 
designs have been proposed, including hybrid systems consisting of ferromagnet and superconductor 
constituents~\cite{NM2013,SuSTMawatari,APL102,SuST26,NJP13,Navau2009} which were recently experimentally realized 
in the forms of multilayered~\cite{AdvMater2012} or bilayered~\cite{Science2012,NJP15} magnet/superconductor hollow cylinder. 
An essential component of the proposed hybrid cylindrical designs is the inner superconducting layer 
which was assumed until quite recently to be an ideal diamagnetic medium with zero effective permeability in both 
analytical and finite-element considerations~\cite{NJP13,Science2012}. 
This assumption is, however, unrealistic because of (1) finite field penetration depth which can be comparable with 
superconductor thickness and (2) massive magnetic flux penetration followed by formation of the critical state typical of magnetic 
shielding applications\cite{Willis,Pavese,Rabbers}.

The system with the finite penetration depth of magnetic field into a superconductor never completely protects the inner region 
(a central hole) from the penetration of weak external field even in the Meissner state~\cite{APL2014}.
On the other hand, a non-distorted uniform 
magnetic field outside the cloak can exist in a wide range of relative permeability and thickness values of the paramagnet 
sheath for both cases of thick and thin superconductor layers being in the Meissner state. At the same time, the
magnetic moment of such a bilayer tube vanishes under the cloaking conditions (as well as all higher multipole moments) making this
object magnetically undetectable. Moreover, penetration of magnetic flux into a superconductor in 
the form of single vortices produces rather small paramagnetic moment of the system, thus breaking 
the perfect cloaking only slightly.

In this respect a following question arises: Is cloaking possible for the case of
multiple vortex penetration at higher fields, and particularly when a superconductor is in the critical
state as is characteristic of shielding with high-temperature superconductors?~\cite{Willis,Pavese,Rabbers} 
In general, comprehensive numerical simulations are necessary to analyze the critical state in structures of such 
kind, even in simpler geometries like a superconductor filament exposed to transverse magnetic 
field~\cite{Ashkin,bookGurevich,CarrJr,Ruiz1,Ruiz2,JPCS2006}. 
However, in the specific case of the superconductor constituent completely 
penetrated by the magnetic flux, {\it i.e.} being in the full critical state, the arising problem is analytically solvable. 

In the present study, based on the Bean concept of the critical state~\cite{Bean}, we demonstrate by exact solving the Maxwell equations for respective media 
that the cloaking effect still holds in a realistic cylindrical design of 
bilayer paramagnet/superconductor tube with finite thicknesses of both superconducting and magnetic 
constituents, when the superconducting layer is already in the full critical state. 
We establish the values of constituent parameters necessary for perfect cloaking and prove also the 
completely vanishing detectability of this object under the cloaking conditions.

\begin{figure}[!tbp]
\includegraphics[width=7.5cm]{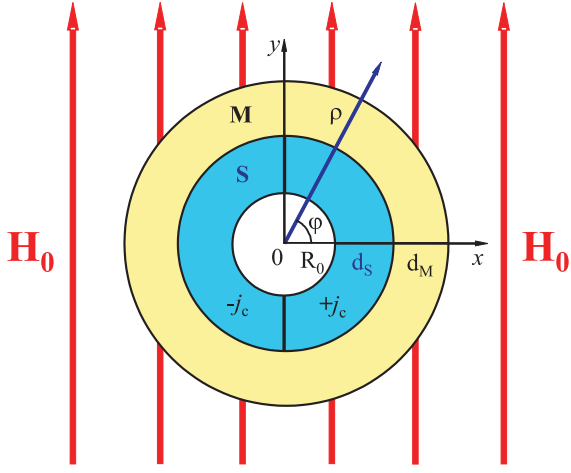}
\caption{(color online) Cross-sectional view of a hollow superconductor cylinder in the full critical state, covered
by a coaxial cylindrical magnetic sheath and assumed to be in the cloaking regime. The vertical solid lines $x=0$ denote 
the boundaries between the superconductor 
regions with negative and positive directions of the critical current. 
The direction of the applied magnetic field $\mathbf{H}_{0}$ is also indicated.} \label{fig1}
\end{figure}

Let us consider an infinitely long hollow superconducting cylinder of thickness $d_S$ and radius of a 
coaxial hole $R_0$ enveloped in a coaxial cylindrical magnetic sheath of thickness $d_M$ with relative 
permeability $\mu > 1$. This structure is exposed to an external constant magnetic field $\mathbf{H}_{0}$ 
perpendicular to the cylinder axis $z$ (see Fig.~\ref{fig1}). 
Magnetic flux enters the superconducting constituent through the
superconductor/magnet interface and induces shielding currents in the regions of flux penetration. According to the
Bean concept of the critical state, the flux-penetrated regions carry a current of density $j_c$ directed
parallel (antiparallel) to the $z$-direction. Generally, these regions evolve in a rather complicated 
manner~\cite{Ashkin,Kuzovlev,JPCS2006,bookGurevich,CarrJr} but, when magnetic flux completely penetrates the superconductor,
the boundary between the regions of positive and negative induced currents
coincides with the $x=0$ plane of Cartesian coordinate system, as marked in Fig.~\ref{fig1}.

In this case, the magnetic field in the system, denoted by $\mathbf{H}_{in}$ inside the hole, by $\mathbf{H}_{S}$ in a superconductor,    
 by $\mathbf{H}_{M}$ in a magnetic sheath and by $\mathbf{H}_{out}$ in a surrounding free space, obey the Maxwell 
equations~\cite{CarrJr}
\begin{equation}
\text{curl } \mathbf{H} = \mathbf{j}_{c} \left[ \theta \left( \rho -R_0 \right) - \theta \left( \rho -R_1 \right) \right], \quad
\text{div } \mathbf{B} =  0,
\label{1}
\end{equation}
with $\mathbf{j}_{c}=  j_c \mbox{ sgn} \left(x \right) \mathbf{e}_z$, where $ \mathbf{e}_z $ is a unit vector in the $z$-direction, 
$\mbox{sgn} (\dots )$ denotes the sign function, $\theta (\dots )$ denotes the Heaviside unit step function,
and $R_1 = R_0 + d_S $, cylindrical coordinates ($\rho ,\varphi ,z$) coaxial with the tube being introduced here for convenience.  
Implying an insulating, nonmagnetic layer of thickness much less than $d_M$ and $d_S$ between the
superconductor and the magnet sheath, which is typical for hybrid magnet/superconductor 
structures (see, for example, Refs.~\cite{Lange2002,Dou2004}), the boundary conditions read~\cite{APL2014}
\begin{subequations}
\label{BC}
\begin{align}
B_{S,n}& =\mu _{0} H_{in,n},\qquad  B_{S,t} =\mu _{0}H_{in,t}; 
\label{3a} \\
B_{S,n}& =\mu _{0}\mu H_{M,n}, \quad B_{S,t} =\mu _{0}H_{M,t}; 
\label{3b} \\
\mu H_{M,n}& =H_{out,n},\quad \quad H_{M,t} =H_{out,t},   \label{3c}
\end{align}
\end{subequations}
for the normal ($n$) and tangential ($t$) components on the inner superconductor 
surface [Eq.~(\ref{3a})], on the superconductor/magnet interface [Eq.~(\ref{3b})] 
and on the outer magnet surface [Eq.~(\ref{3c})], respectively (cf. also Refs.~\cite{APL2004,PRB2005}), 
with $\mu_0$ permeability of vacuum. In addition, the magnetic field outside the structure has to approach asymptotically the 
applied field $\mathbf{H}_{0}$.

In cylindrical coordinates ($\rho ,\varphi ,z$), the solution of Eqs.~(\ref{1}) takes the form:
\begin{eqnarray}
H_{in,\rho} &=&A_{in}\sin \varphi, \nonumber \label{Hhole} \\
H_{in,\varphi} &=&A_{in} \cos \varphi,   
\end{eqnarray}
in the hole ($ \rho \le R_0$);
\begin{eqnarray}
B_{S,\rho} & = & \mu_0 \left( A_{S1}/\rho^2 + A_{S2} + 4 j_c \rho /3 \pi  \right)  \sin \varphi, \nonumber \label{HSC} \\
B_{S,\varphi} & = & \mu_0 \left( -A_{S1}/\rho^2 + A_{S2} + 8 j_c \rho /3 \pi \right)  \cos \varphi, 
\end{eqnarray}
in the superconductor ($ R_0 \le \rho \le R_1$);
\begin{eqnarray}
H_{M,\rho} &=&\left( A_{M1}/\rho^2 + A_{M2} \right) \sin \varphi,  \label{HML} \nonumber\\
H_{M,\varphi} &=&\left( -A_{M1}/\rho^2 + A_{M2} \right) \cos \varphi,   
\end{eqnarray}%
in the magnet sheath ($ R_1 \le \rho \le R_2= R_1 + d_M$); and
\begin{eqnarray}
H_{out,\rho} &=& \left( H_{0}+A_{out}/\rho^2\right) \sin \varphi,   \label{Hspace} \nonumber \\
H_{out,\varphi} &=&\left( H_{0}-A_{out}/\rho^2\right) \cos \varphi,   
\end{eqnarray}%
in the space around the tube ($ \rho \ge R_2$). 
The coefficients $A_{in}$, $A_{S1}$, $A_{S2}$, $A_{M1}$, $A_{M2}$ and 
$A_{out}$ determined from the boundary conditions~(\ref{BC}) are given by the following expressions
\begin{eqnarray}
A_{in} &=& 4 \mu R_2^2 H_0 /\Delta - \left( 2 j_c /3 \pi \right)  \left[ 3 \left( R_1 - R_0 \right)  \right.  \nonumber \\
&+&  \left. \left( \mu^2 - 1 \right) \left(R_2^2 - R_1^2 \right) \left(R_1^3 - R_0^3 \right) / R_1^2 \Delta \right] , \label{Ain} \\
A_{S1} &=& 2 j_c R_0^3 / 3 \pi, \\
A_{S2} &=& 4 \mu R_2^2 H_0 /\Delta - \left( 2 j_c /3 \pi \right) \left[ 3 R_1 + \left( \mu^2 - 1 \right) \right.  \nonumber \\
&\times&  \left.  \left(R_2^2 - R_1^2 \right) \left(R_1^3 - R_0^3 \right) /R_1^2 \Delta \right], \\
A_{M1} &=& -2 H_0 R_1^2 R_2^2 \left(\mu - 1 \right)/\Delta  \nonumber \\
&-& \left(4 j_c/ 3 \pi\right) \left(\mu + 1 \right) \left(R_1^3 - R_0^3 \right) R_2^2 /\Delta, \label{AM1} \\
A_{M2} &=& 2 H_0 R_2^2 \left(\mu +1 \right)/ \Delta \nonumber \\
&+& \left(4 j_c /3 \pi \right) \left(\mu - 1 \right) \left(R_1^3 - R_0^3 \right)/\Delta, \label{AM2}\\
A_{out} &=& H_0 R_2^2 \left(\mu^2 - 1 \right) \left(R_2^2-R_1^2 \right)/ \Delta \nonumber \\
&-& \left(8 j_c/ 3 \pi\right) \mu \left( R_1^3 - R_0^3 \right)/\Delta, \label{Aout}
\end{eqnarray}
with $ \Delta = \left( \mu + 1\right)^2 R_2^2 - \left( \mu - 1\right)^2 R_1^2$.

The obtained expressions~(\ref{Ain})-(\ref{Aout}) are valid only for  
applied fields $H_0$ equal or higher than the field of full flux penetration in the superconducting constituent, $H_{fp}$. 
The latter is determined by the condition
\begin{equation}
B_S|_{\rho = R_0} = \mu_0 \left(A_{S1}/R_0^2 + A_{S2} + 4 j_c R_0/3 \pi \right) = 0,
\end{equation}  
and reads
\begin{eqnarray}
H_{fp} &=& H_{fp}^0 \left[ 1 + \frac{\left(\mu -1 \right) \left(2 \mu -1 \right)}{6 \mu} \frac{\left(R_2^2- R_1^2 \right)}{R_2^2}\right. \nonumber \\
 &+& \left. \frac{\left(\mu^2 -1 \right)}{12 \mu} \frac{\left(R_2^2 - R_1^2 \right) R_0 \left(R_0 + R_1 \right)}{ R_1^2 R_2^2} \right],  \label{Hp}
\end{eqnarray}
where $H_{fp}^0 = (2 j_c / \pi) (R_1 -R_0)$ is the field of full flux penetration in the superconductor constituent 
for an unshielded superconductor cylinder~\cite{CarrJr}.
In the limit of $R_0 \to 0$, $H_{fp}$ reduces to the known expression for the magnetically shielded superconductor filament~\cite{JPCS2006}. 

\begin{figure}[!tbp]
\includegraphics[width=8.5cm]{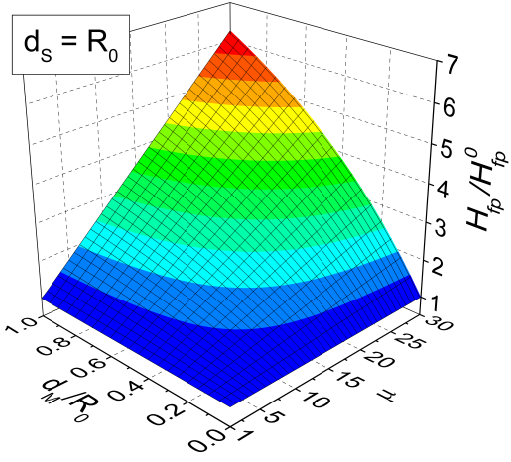}
\caption{(color online) The dependence of the field of the full flux penetration $H_{fp}$ on the relative permeability $\mu$ and on the thickness $d_M$ of the
magnet sheath for the thickness of superconductor layer $d_S = R_0$.}
\label{figHp1}
\end{figure}
In Fig.~\ref{figHp1} we present a typical dependence of the field $H_{fp}$ on the relative permeability $\mu$ and on the thickness $d_M$ 
of the magnet sheath calculated exemplarily for the ratio of superconductor thickness  to the inner hole radius $d_S/ R_0 = 1$. One can see that 
due to the shielding of the applied field by the magnetic sheath the field of the full flux penetration generally increases.
With the increase of $\mu$ at a fixed thickness $d_M$
the $H_{fp}(d_M, \mu ) $ dependence approximates to the linear one. A monotonic behavior is also demonstrated by the $H_{fp}(d_M, \mu ) $ dependence
with the increase of $d_M$ at a fixed permeability $\mu$. For other values of the ratio $d_S/ R_0$ the $ H_{fp}(d_M, \mu ) $ dependences reveal a 
qualitatively similar behavior and differ only by the scale of the magnitude. In the limiting case of $d_S \ll R_0 $ this field is reduced to
\begin{equation}
H_{fp} = H_{fp}^0 \left[ 1 + \frac{\left(\mu -1 \right)}{2} \frac{d_M \left(2  R_0 + d_M \right)}{\left( R_0 + d_M \right)^2}\right].
\end{equation}
Considering the opposite limit $ d_S \gg R_0$, closed to the case of the magnetically shielded superconducting filament~\cite{JPCS2006}, we obtain
\begin{equation}
H_{fp} = H_{fp}^0 \left[ 1 + \frac{\left(\mu -1 \right) \left(2 \mu -1 \right)}{6 \mu} \frac{d_M}{d_S} \right].
\end{equation}

In order to retain the magnetic field outside the cloak undisturbed, a condition $A_{out} = 0$ has to be fulfilled.
This results in the following equation:
\begin{equation}
\frac{H_{cl}}{H_{fp}^0} = \frac{4}{3} \frac{\mu}{\left( \mu^2 -1 \right)} \frac{R_1^2 + R_0 R_1 + R_0^2}{R_2^2 - R_1^2},  \label{CloakExt}
\end{equation}
from which, for given fixed values of parameters $\mu, d_M $ and $d_S$, a magnitude of applied magnetic field $ H_{cl} \ge H_{fp}$ providing 
the cloaking effect can be found. Notice that this is qualitatively different from the case of cloaking in the Meissner state, 
where the cloaking conditions were field-independent~\cite{APL2014}. 

\begin{figure}[!tbp]
\includegraphics[width=8.5cm]{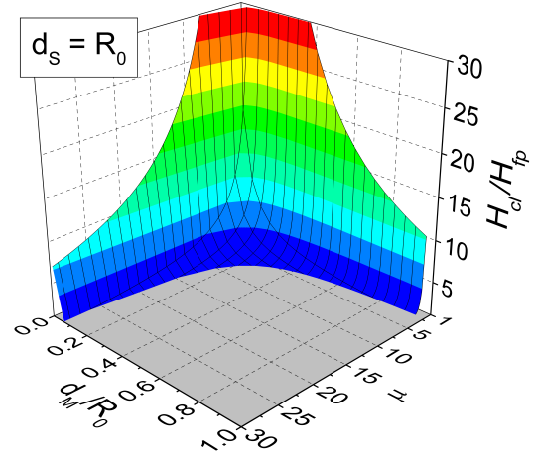}
\caption{(color online) The dependence of the ratio $H_{cl} / H_{fp}$ on the relative permeability $\mu$ and on the thickness $d_M$ of the
magnet sheath for the thickness of superconductor layer $d_S = R_0$.} 
\label{figHcl}
\end{figure}

The calculated dependence of $ H_{cl}/H_{fp}$ on the relative permeability $\mu$ and on the thickness $d_M$ is shown in Fig.~\ref{figHcl} 
for the same value of ratio of $d_S  /R_0$ as in Fig.~\ref{figHp1}.
\begin{figure}[!bp]
\includegraphics[width=8cm]{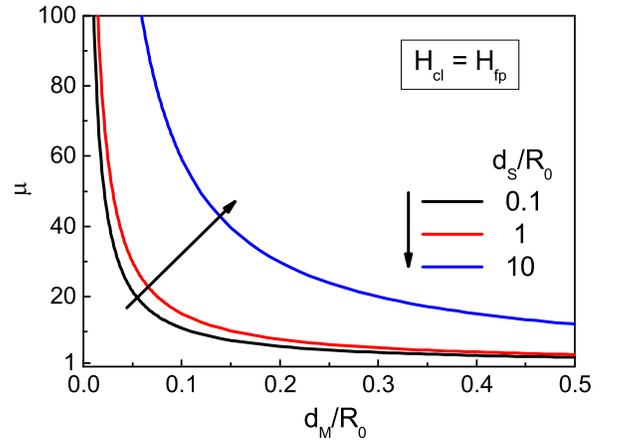}
\caption{(color online) Solutions of Eq.~(\ref{CloakExt}) with the magnetic field $H_{cl} = H_{fp}$ for different values of the superconductor layer thickness $d_S$.
} 
\label{fig2}
\end{figure}
One can see that the cloaking effect exists for a sufficiently wide range of parameters of the magnet layer restricted by the $\mu (d_M) $ curve 
corresponding to the solution of Eq.~(\ref{CloakExt}) at $H_{cl} = H_{fp}$. This boundary is shown in Fig.~\ref{fig2} for the different values of relative 
superconductor thickness $d_S/R_0$. 
The $H_{cl} (\mu, d_M)$ surfaces for other $d_S/R_0$ values are qualitatively similar to that of Fig.~\ref{figHcl} differing 
only by the scale of the magnitude. In the case of rather thin ($d_M \ll R_1 $) magnetic sheath 
the cloaking effect can be provided for both thin ($d_S \ll R_0$) and thick ($d_S \gg R_0$) superconductors if the relative permeability satisfies the condition
$\mu d_M/R_1\simeq H_{fp}^0 / H_0$. This is similar to the known effectiveness of magnetic shielding for the Meissner state of the superconducting constituent 
as soon as the strength of an effective magnetic dipole layer $\mu d_M/R_1$ is notable~\cite{GenenkoJAP2002,GenenkoPhysC2004-1}.   

It's interesting to note that also the $\mu (d_M) $ curves in Fig.~\ref{fig2} are qualitatively similar to those obtained earlier~\cite{APL2014} for 
cloaking in the case of superconducting constituent being in the Meissner state (cf. Fig.~2 of Ref.~\cite{APL2014}). This hints that the specific 
cloaking parameters of the structure calculated for both cases -- of the Meissner state and of the full critical state -- can be related each other.
Indeed, using in Eq.~(\ref{CloakExt}) the cloaking parameters corresponding to the former case, one can calculate the field $H_{cl}$. 
If the obtained value will satisfy the condition $H_{cl} \ge H_{fp}$, then, having in mind possible breaking of cloaking at fields 
when the magnetic flux starts to penetrate the superconductor~\cite{APL2014}, it would mean that at the field $H_0 = H_{cl}$ cloaking appears again.

As for the possible screening of the inner space of the cloak from the external magnetic field,
by using Eq.~(\ref{Hp}), it follows that 
\begin{equation}
A_{in} = \left( 4 \mu R_2^2/\Delta\right) \left( H_0 - H_{fp} \right), \label{AinHp}
\end{equation}
and, therefore, at field $H_0 = H_{fp}$ the magnetic field inside the cloak vanishes revealing complete protection of the inner hole of the structure
from the penetration of magnetic flux. At higher applied fields, a non-zero homogeneous magnetic field 
$H_{y,in} = A_{in}$ always exists inside the cloak though reduced with respect to the external value.

Another important question concerning the "invisibility" of the coaxial bilayer structure is whether or not 
it can be detected~\cite{NJP13}, for example, by measuring a magnetic moment of this structure. 
Due to the geometry of the system, this moment per unit length of the structure has only a $y$ component and consists of two parts 
(cf. Refs.~\cite{APL2004,APL2014}): the magnetic moment of the superconductor layer defined by means of the critical current density 
$\mathbf{j}_c$ as 
\begin{equation}
M_S =  \int_{V_S} dV \left[ \bm{\rho} \times \mathbf{j}_c \right]_y,  
\end{equation}
and the moment of magnetic sheath defined as
\begin{equation}
M_M =  \left( \mu -1 \right) \int_{V_M} dV H_{M,y}.
\end{equation}
Using Eqs.~(\ref{HML}), (\ref{AM1}) and (\ref{AM2}), after integration one obtains
\begin{eqnarray}
M_S &=& -\left( 4 j_c / 3\right)\left(R_1^3 - R_0^3 \right), \\
M_M &=&  \pi \left(\mu -1 \right) A_{M2} \left(R_2^2 - R_1^2 \right),
\end{eqnarray}
and it is easy to ensure that the total magnetic moment $M=M_S+M_M$ per unit length of the paramagnet/superconductor tube being in the 
superconducting full critical state is expressed through the coefficient $A_{out}$,
\begin{equation}
M = 2 \pi A_{out}.
\end{equation}
This means that in the cloaking case ($A_{out}=0$, i.e. $H_0 = H_{cl}$) the magnetic moment of the structure under consideration simultaneously vanishes 
ensuring that the object cannot be detected by magnetic measurements. This is because, according to the form of the solution~(\ref{Hhole})-(\ref{Hspace}), 
the structure does not possess other multipole moments but the dipole one. If the latter equals zero the object cannot be observed by any magnetic
measurement, at least as long as an external field uniform at the scale of the object transverse size 
$\sim R_2$ is involved. Thus, magnetic undetectability of the system happening in the Meissner state can be reestablished at the field $H_{cl}$.

In conclusion, we have studied theoretically static magnetic cloaking properties of a realistic bilayer 
paramagnet/superconductor cylindrical tube in the case when the superconducting constituent is completely 
penetrated by the applied transverse magnetic field. We have found that a non-distorted uniform magnetic field 
outside the cloak can exist in a wide range of relative permeability and thickness values of the magnet 
sheath for both cases of thick and thin superconductor layers but only at the specific value of applied magnetic field. 
Under the above cloaking conditions the
magnetic moment of the bilayer structure vanishes (as well as all higher multipole moments) making this
object magnetically undetectable. When the applied field equals to the field of full flux penetration, 
such a system also completely protects the inner region (a central hole) from the penetration of the external field, 
thus revealing in this case cloaking in its ideal, "dual" form. At higher applied fields, the central hole of the 
structure is never protected from the magnetic flux penetration. These results are expected to hold also for low frequency ac fields as is 
generally the case for superconducting shielding~\cite{Willis,Pavese,Rabbers}.

Thus, we have demonstrated that, along with a manifestation of both cloaking and complete magnetic undetectability 
by a paramagnet/superconductor cylindrical structure 
in the Meissner state known in the literature~\cite{NJP13,Science2012,NJP15,APL2014,SuST26}, 
these features can be reestablished by increase of an applied magnetic field to a certain value in the range of the superconductor full critical state. 
The question, whether cloaking (and magnetic undetectability) is possible when the superconductor constituent is only partly 
penetrated by the magnetic flux, even in the form of a few vortices, needs however a more elaborated treatment. 
   
\bibliographystyle{plain}
\bibliography{apssamp}

\begin{thebibliography}{99}

\bibitem{review2012} N.I.~Zheludev and Y.S.~Kivshar, Nature Mater. {\bf 11}, 917 (2012).

\bibitem{Science1} U.~Leonhardt, Science {\bf 312}, 1777 (2006).

\bibitem{Science2} J.B.~Pendry, D.~Schurig, and D.R.~Smith, Science {\bf 312}, 1780 (2006).

\bibitem{Science3} D.~Schurig, J.J. Mock, B.J. Justice, S.A. Cummer, J.B.~Pendry, A.F. Starr, 
and D.R.~Smith, Science {\bf 314}, 977 (2006).

\bibitem{Pendry2007} B.~Wood and J.B.~Pendry, J.~Phys.: Condens. Matter {\bf 19}, 076208 (2007).

\bibitem{NM2008} F.~Magnus, B.~Wood, J.~Moore, K.~Morrison, G.~Perkins, J.~Fyson, M.C.K.~Wiltshire, 
D.~Caplin, L.F.~Cohen, and J.B.~Pendry, Nature Mater. {\bf 7}, 295 (2008).

\bibitem{ElCloak} F.~Yang, Z.L.~Mei, T.Y.~Jin, and T.J.~Cui, Phys. Rev. Lett. {\bf 109}, 053902 (2012).

\bibitem{NM2013} N.~Landy and D.R.~Smith, Nature Mater. {\bf 12}, 25 (2013).

\bibitem{Navau2009} C.~Navau, D.X.~Chen, A.~Sanchez, and N.~Del-Valle, Appl. Phys. Lett. {\bf 94}, 242501 (2009).

\bibitem{NJP13} A.~Sanchez, C.~Navau, J.~Prat-Camps, and D.X.~Chen, New J. Phys. {\bf 13}, 093034 (2011).

\bibitem{SuST26} J.~Prat-Camps, A.~Sanchez, and C.~Navau, Supercond. Sci. Technol. {\bf 26}, 074001 (2013).

\bibitem{SuSTMawatari} Y.~Mawatari, Supercond. Sci. Technol. {\bf 26}, 074005 (2013).

\bibitem{APL102} R.~Wang, Z.L.~Mei, and T.J.~Cui, Appl. Phys. Lett. {\bf 102}, 213501 (2013).

\bibitem{AdvMater2012} S.~Narayana and Y.~Sato, Adv. Mater. {\bf 24}, 71 (2012).

\bibitem{Science2012} F.~G\"om\"ory, M.~Solovyov, J.~\v{S}ouc, C.~Navau, J.~Prat-Camps, and A.~Sanchez, 
Science {\bf 335}, 1466 (2012).

\bibitem{NJP15} J.~\v{S}ouc, M.~Solovyov, F.~G\"om\"ory, J.~Prat-Camps, C.~Navau, and A.~Sanchez,
New J. Phys. {\bf 15}, 053019 (2013).

\bibitem{Willis} J.O.~Willis, M.E.~McHenry, M.P.~Maley, and H.~Sheinberg, IEEE Trans. Magn. {\bf 25}, 2502 (1989).

\bibitem{Pavese} F. Pavese, {\it Magnetic shielding}, in {\it Handbook of Applied Superconductivity} (IoP Publishing, Bristol, UK, 1998), p.~1461-1483.

\bibitem{Rabbers} J.J.~Rabbers, M.P.~Oomen, E.~Bassani, G.~Ripamonti, and G.~Giunchi, Supercond. Sci. Technol. {\bf 23}, 125003 (2010).

\bibitem{APL2014} S.V.~Yampolskii and Y.A.~Genenko, Appl. Phys. Lett. {\bf 104}, 033501 (2014).

\bibitem{Ashkin} M.~Ashkin, J. Appl. Phys. {\bf 50}, 7060 (1979).

\bibitem{bookGurevich} A.V.~Gurevich, R.G.~Mints, and A.L.~Rakhmanov, {\it The Physics of Composite
Superconductors} (Begell House, New York, 1997).

\bibitem{CarrJr} W.J.~Carr, Jr., \textit{AC Loss and Macroscopic Theory of Superconductors}, 2nd ed. (Taylor \& Francis, London, 2001).

\bibitem{JPCS2006} S.V.~Yampolskii, Y.A.~Genenko, H.~Rauh, and A.V.~Snezhko, J.~Phys.: Conf. Ser. {\bf 43}, 576 (2006).

\bibitem{Ruiz1} H.S.~Ruiz, A.~Bad\'{\i}a-Maj\'{o}s, Y.A.~Genenko, H.~Rauh, and S.V.~Yampolskii, Appl. 
Phys. Lett. {\bf 100}, 112602 (2012).

\bibitem{Ruiz2} H.S.~Ruiz and A.~Bad\'{\i}a-Maj\'{o}s, J. Appl. Phys. {\bf 113}, 193906 (2013).

\bibitem{Bean} C.P.~Bean, Phys. Rev. Lett. {\bf 8}, 250 (1962).

\bibitem{Kuzovlev} Y. E. Kuzovlev, Pis'ma Zh. Eksp. Teor. Fiz. {\bf 61}, 970 (1995) [JETP Lett. {\bf 61}, 1000 (1995)].

\bibitem{Lange2002} M.~Lange, M.J.~Van Bael, V.V.~Moshchalkov, and Y.~Bruynseraede, Appl. Phys. Lett.
{\bf 81}, 322 (2002).

\bibitem{Dou2004} A.V.~Pan and S.X.~Dou, J.~Appl. Phys. {\bf 96}, 1146 (2004).

\bibitem{APL2004} Y.A.~Genenko, S.V.~Yampolskii, and A.V.~Pan, Appl. Phys. Lett. {\bf 84}, 3921 (2004).

\bibitem{PRB2005} S.V.~Yampolskii and Y.A.~Genenko, Phys. Rev. B {\bf 71}, 134519 (2005).

\bibitem{GenenkoJAP2002} Y.A.~Genenko and A.V.~Snezhko, J.~Appl. Phys. {\bf 92}, 357 (2002). 

\bibitem{GenenkoPhysC2004-1} Y.A.~Genenko, A.V.~Snezhko, and A.~Usoskin, Physica C {\bf 401}, 236 (2004).


\end{thebibliography}

\end{document}